\documentclass[aps,prl,twocolumn,superscriptaddress]{revtex4}

\usepackage[draft]{hyperref}
\usepackage{amsmath,amssymb}
\usepackage{graphicx}

\newcommand{\be}{\begin{equation}}
\newcommand{\ee}{\end{equation}}
\newcommand{\ba}{\begin{eqnarray}}
\newcommand{\ea}{\end{eqnarray}}
\newcommand{\bra}[1]{\ensuremath{\langle #1 |}}
\newcommand{\ket}[1]{\ensuremath{|#1\rangle}}

\newcommand{\braket}[1]{\ensuremath{\langle #1\rangle}}

\begin{document}

\title{Probing the negative Wigner function of a pulsed single photon point by point} 


\author{Kaisa Laiho}
 \email{Kaisa.Laiho@mpl.mpg.de}
\affiliation{Max Planck Institute for the Science of Light, G\"{u}nther-Scharowsky-stra\ss e 1/Building 24, 91058 Erlangen, Germany.}
\author{Kati\'uscia N. Cassemiro}
\affiliation{Max Planck Institute for the Science of Light, G\"{u}nther-Scharowsky-stra\ss e 1/Building 24, 91058 Erlangen, Germany.}
\author{David Gross}
\affiliation{Institute for Theoretical Physics, ETH Z\"{u}rich, 8093 Z\"{u}rich, Switzerland}
\author{Christine Silberhorn}
\affiliation{Max Planck Institute for the Science of Light, G\"{u}nther-Scharowsky-stra\ss e 1/Building 24, 91058 Erlangen, Germany.}
\affiliation{Applied Physics, University of Paderborn,  Warburgerstra\ss e 100, 33098 Paderborn, Germany.}

\date{\today}

\begin{abstract}
We investigate quantum properties of pulsed light fields point by point in phase space. We probe the negative region of the Wigner function of a single photon generated by the means of  waveguided parametric down-conversion.  This capability is achieved by employing loss-tolerant photon-number resolving detection, allowing us to directly observe the oscillations of the photon statistics in dependence of applied displacements in phase space. Our scheme is highly mode sensitive and can reveal the single-mode character of the signal state.
\end{abstract}


\maketitle


Quantum tomography constitutes an essential tool for the exploration and verification of quantum information tasks. For continuous variables (CV) systems,  homodyne detection is an ubiquitous technique allowing the characterization of quantum states and processes by the Wigner function in terms of the field quadratures~\cite{Lvovsky2009}.  The Wigner representation 
is  particularly expedient for studying genuine quantum properties  of coherent light fields or the nonclassicality of non-Gaussian states, which is signalized by the negative values of the distribution~\cite{A.I.Lvovsky2001, Luetkenhaus95}.
However, as the Heisenberg's uncertainty principle precludes the simultaneous measurement of 
a pair of noncommuting quadratures, the evaluation of the Wigner function at a specific point in phase space by homodyne detection is only possible by means of tomographic reconstruction.

Surprisingly, we can nevertheless  directly access the value of the Wigner function at a single point in phase space by changing the detection method; this means the measured observable~\cite{Englert1993}.   This technique, which was first demonstrated with motional quantum states of a trapped ion~\cite{Leibfried1996}, requires the measurement of only  \emph{one} operator---namely, the parity.  In optical experiments this can be realized using photon counting~\cite{K.Banaszek1996}.
 Although conceptually simple, up to now the use of this technique, referred to as \emph{direct probing}, has  been limited by the lack of optical photon number resolving (PNR) detectors. It has been applied to optical states confined to a cavity, for which atom-photon interactions were utilized rather than direct photo-detection~\cite{P.Bertet2002}. For traveling light fields, the first proof-of-principle experiment demonstrated the method for coherent states in the continuous wave regime~\cite{K.Banaszek1999}. Lately, Gaussian states of light have been characterized  in the pulsed regime even without true PNR detectors~\cite{Bondani2009}. 

However, a point-by-point measurement of  a non-Gaussian state exhibiting negative values in the Wigner function with free propagating light pulses has so far not  been reported.  Apart from the fundamental interest, the successful demonstration of such an experiment opens a new path for the state analysis in the context of quantum communication protocols. It can be highly attractive for the study of CV Bell inequalities~\cite{Banaszek1998} or CV entanglement distillation associated with the famous no-go theorem \cite{J.Eisert2002} that revealed the essential need of non-Gaussianity.

One recently established PNR detection technique for pulsed light utilizes time-multiplexing~\cite{D.Achilles2003} enabling loss-tolerant detection of photon statistics~\cite{D.Achilles2005}. 
The time-multiplexed detector  (TMD) has  lately been in the focus of increased attention, as it is the first  detector whose POVMs have been tomographically characterized~\cite{Lundeen2009}.  In a more sophisticated configuration,  it allows the realization of configurable projective measurements~\cite{Puentes2009}. Considering state characterization, a loss-independent measurement of higher order moments can be efficiently accomplished with the TMD network~\cite{Avenhaus2010}. These features, together with the advantages of photon counting for state discrimination~\cite{Wittmann2010}, make the TMD a promising and powerful tool for preparing and analyzing quantum states.

In this Letter we apply the TMD to directly measure the statistics of displaced, pulsed single-photon states and
probe point by point the corresponding Wigner function. We uncover the non-Gaussianity of the
single-photon state by observing the characteristic oscillating behavior of the photon statistics in the dependence of the displacement in phase space.
Our focus lies on the ability to prepare and manipulate  pulsed nonclassical states of light exhibiting broad spectra as well as on combining loss tolerance with mode-sensitive detection.

\begin{figure}[b]
\includegraphics[width = 0.48\textwidth]{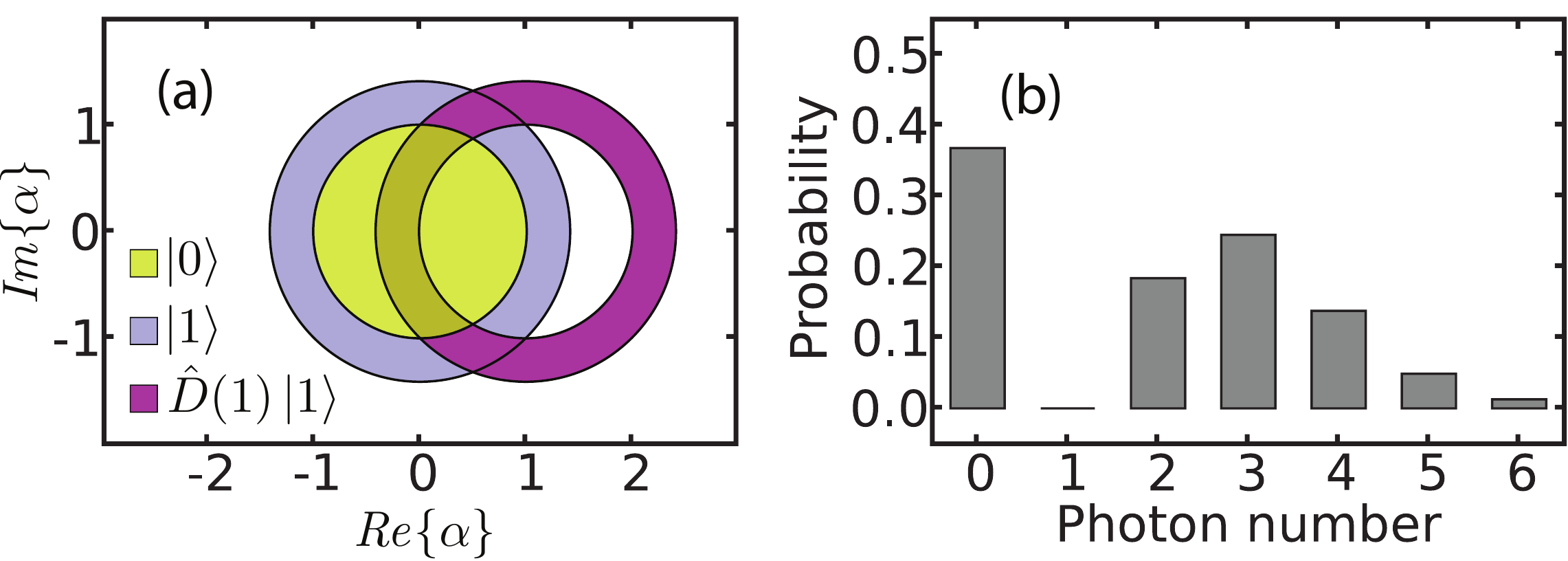}
\caption{Interpretation of photon counting.  (a) The detector projects into number states corresponding to rings in phase space ($\ket{0}$ and $\ket{1}$ are shown).  The displaced state $\hat{D}(1)\ket{1}$ is shifted from the origin by $\alpha=1$. (b) The semiclassical model predicts the photon-number oscillation in the statistics of $\hat{D}(1)\ket{1}$.}
\label{fig_rings}
\end{figure}

The value of the Wigner function at position $\alpha$  is given by the expectation value of the parity operator on the probed quantum state $\varrho$ displaced by $-\alpha$~\cite{K.Banaszek1996} 
\begin{equation}
W(\alpha) = 2/\pi \ \textrm{Tr} \{ \hat{\Pi} \ \hat{D}^{\dagger}(\alpha) \varrho \hat{D}(\alpha)  \}\,,
\label{eq_W}
\end{equation}
with $\hat{\Pi} = (-1)^{\hat{n}}$ being the parity operator, $\hat{n}$ the photon number operator and $\hat{D}(\alpha)$  the displacement. Since the Fock states are parity eigenstates,
the measurement of parity can be realized  by recording photon statistics.
The determination of the Wigner function by photon counting  significantly differs from homodyne tomography. Homodyne detection projects the quantum state in the phase space along a one-dimensional field quadrature and  constitutes a Gaussian measurement. A PNR detector, on the other hand, realizes a projection into the photon-number basis $\{ \ket{n}\}$, which corresponds to a non-Gaussian operation.
Heuristically, the photon number measurements can be understood as projections onto rings, the radius of which are determined by the excitation number $n$ [Fig.~\ref{fig_rings}(a)]. The displacement operation in Eq.~(\ref{eq_W}) shifts the quantum  state away from the origin, which changes its overlap with the number states. By considering the intersections of the phase-space diagrams describing the displaced state and the detection, one can interpret the resulting statistics with a semiclassical area-of-overlap principle ~\cite{W.Schleich1987}.
The semiclassical model does not only take into account the intersection areas, but 
also crucially relies on the interference of two separate regions  in phase space [diamond shape in Fig.~\ref{fig_rings}(a)] that give rise to a characteristic oscillation of the photon number distribution [Fig.~\ref{fig_rings}(b)]~\cite{Oliveira1990}. Thus, these measurements also probe the coherence of the entire signal state in phase space. 

Another important difference between direct probing and homodyne detection is the fact that in the latter an intrinsic filtering operation takes place. In order to measure the quantum noise with conventional photodiodes, homodyne detection  relies on the interference of a strong local oscillator field with a weak quantum signal. Only the part of the signal state that overlaps with the local oscillator can be seen by the detector and is ``amplified.'' Thus, loss and mode mismatch give the same signature. 
On the other hand, the method of direct probing detects all modes, resulting in a more complete state characterization with an intrinsic quantification of the mode overlap~\cite{Banaszek2002, K.Laiho2009}.
The displacement, needed to probe the Wigner function at an arbitrary point of the phase space, is experimentally accomplished by the use of an asymmetric beam splitter. The signal is  overlapped with a weak single-mode coherent reference beam, yielding at a displacement of $-\alpha$  the corresponding  photon-number distribution $\rho_{n}$. The mode mismatch between those fields leads to a convolution of a Poissonian term with mean photon number of $(1-\mathcal{M})|\alpha|^{2}$ and  a displaced  part
\begin{equation}
 \rho_{n}^D= \textrm{Tr}\{  \ket{n}\bra{n} \ \ \hat{D}^{\dagger}(\sqrt{\mathcal{M}}\alpha) \varrho  \hat{D}(\sqrt{\mathcal{M}}\alpha ) \}\,,
 \label{eq_overlap}
 \end{equation}
in which the amount of overlap is quantified by a parameter $0 \le\mathcal{M}\le1$.

 The TMD, employed to measure the photon statistics, 
consists  of  a network of symmetric (50/50) beam splitters with different lengths of fiber loops in between. A pulse is divided in the network into two pulse trains that are subsequently detected by two avalanche photo diodes (APDs). 
As described in~\cite{D.Achilles2004}, the TMD is characterized by the loss  and convolution  matrices, $\mathbf{L}(\eta)$ and $\mathbf{C}$ respectively. The former depends on the detection efficiency $\eta$, whereas the latter accounts  for the stochastic splitting of the photons at the beam splitter stages. In an ensemble measurement, the TMD records click statistics  $(\vec{p}_{click} = [p_{0}, p_{1}, \dots p_{n}]^{T})$, which are related to the photon statistics of the state  $(\vec{\rho} = [\rho_{0}, \rho_{1}, \dots \rho_{n}]^{T})$ by the simple expression $\vec{p}_{click} = \mathbf{CL}({\eta})\vec{\rho}$. Loss-tolerant detection is achieved by inverting this matrix relation \cite{D.Achilles2005}, and the mean value of the parity can be extracted from the inverted statistics by the alternating sum $\braket{\hat{\Pi}}= \sum_{n} (-1)^{n}\rho_{n}$.

\begin{figure}[b]
\includegraphics[width = 0.48\textwidth]{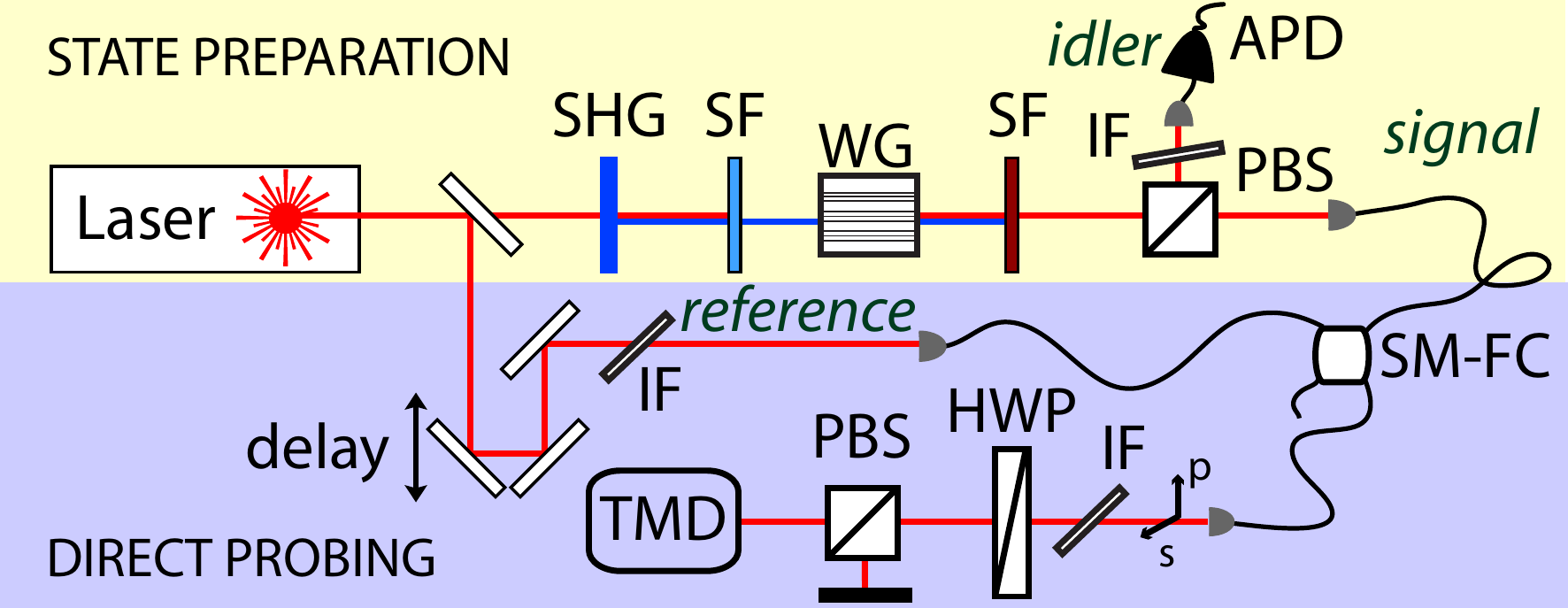}
\caption{Experimental setup. SHG, second harmonic generation in thin nonlinear crystal; SF, spectral filter for blocking the residual pump. For further details see text.}
\label{fig_setup}
\end{figure} 

To demonstrate the capabilities of the direct probing method, we choose to study a single-photon state with a complicated mode structure in the spectral degree of freedom, but possessing only diagonal elements in the density matrix. This simplifies the detection scheme, precluding the need of phase sensitivity. As shown in Fig.~\ref{fig_setup}, we prepared our single-photon state by heralding on one of the twin beams produced by type-II parametric down-conversion (PDC) in a 1.45~mm long, periodically poled $\textrm{KTiOPO}_{4}$ waveguide (WG). This was pumped by a frequency doubled  Ti:Sapphire laser (wavelength 796~nm, bandwidth 10~nm, repetition rate 2~MHz). After the separation of the orthogonally polarized signal and idler twin beams by  a polarizing beam splitter (PBS), we launched both of them separately into single-mode (SM) fibers for spatial mode cleaning. The idler was filtered by a 1~nm broad interference filter (IF) to ensure good spectral SM characteristics for the heralding with a single APD. For the implementation of the reference beam a small fraction of the initial laser beam was attenuated to the single-photon level and filtered to a bandwidth of 1~nm.

The proper realization of the displacement operation requires the control of the photonic states in all degrees of freedom: spectral, spatial and temporal. To achieve a good spatial overlap between signal and reference we employed  a SM fiber coupler (SM-FC). This  acts as PBS such that  the horizontal (p) and vertical (s) polarization, referring to signal and reference, were combined at a single output port. In order to suppress the background of simultaneously excited signal modes we added a 3~nm broad filter to the joint beam path. Finally, both beams were sent to a variable ratio beam splitter, constructed of a half-wave plate (HWP) and a PBS, and detected with the TMD.
By measuring the Hong-Ou-Mandel interference between signal and reference as described in \cite{Laiho2009} we analyzed the temporal and spectral overlaps.
This yielded a  maximal overlap of 0.71(4) between signal and reference. 
 
To measure the displaced statistics, the beam splitter  was set to transmit over 90\% of the signal.  The TMD data was collected with a  time-to-digital converter, which records the detection times of the APD clicks with respect   to the repetition time of the laser. Therefore, we could suppress background and dark count events
by applying a tight time gating ($<$4~ns) in each time bin of the detection channels.  By post-processing  the data 
 we constructed the heralded statistics by conditioning on idler click events. Additionally, the same data provided  the probability distribution of the TMD bin populations, needed for calibrating the  $\mathbf{C}$ matrix. The typical accumulation time for collecting statistics with more than $10^{6}$ conditional events  was on the order of $10^{3}$~s. 
 
The full capacity of  our detector is gained after the determination of  the efficiency and the effective displacement. Concerning the former, we take advantage of the PDC that always produces photons in pairs. This allows us to gauge the efficiency by the Klyshko method \cite{D.Achilles2005}  
 in terms of the ratio between the signal-idler coincidence rate and the single-count rate of idler. We measured $\eta = 0.165$, which was  employed 
  throughout the whole data set. We attribute the losses mainly to the limited quantum efficiency of the detectors and to the coupling of   the waveguide spatial mode into SM fibers. The value of the displacement is calibrated  by $|\alpha| = \sqrt{\braket{n}_{\textrm{ref}}/\eta}$, where $\braket{n}_{\textrm{ref}}$ is the mean photon number of the reference measured by blocking the signal arm~\cite{K.Laiho2009}. Thus, all the parameters required for a loss-tolerant Wigner function reconstruction are directly measured. 

\begin{figure}[b]
\includegraphics[width = 0.48\textwidth]{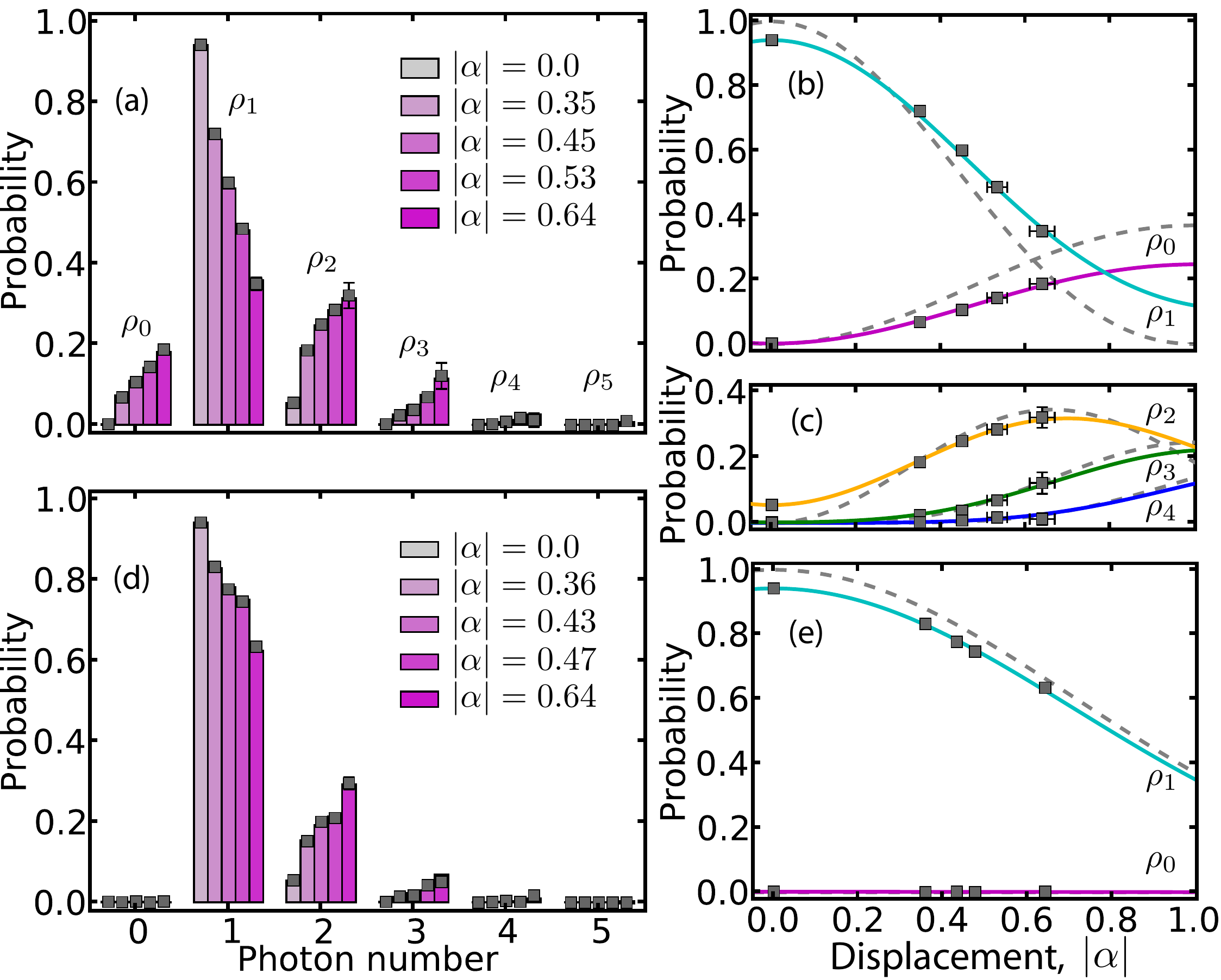}
\caption{(a) Measured statistics for different values of displacement $|\alpha|$ for temporally matched fields---case (I).  Detailed behavior of different photon-number contributions: (b) $\rho_{0}$ and $\rho_{1}$ as well as (c) $\rho_{2}$--$\rho_{4}$ with respect to $|\alpha|$.  From (b) we extract $\mathcal{M}_{\textrm{I}} = 0.70(2)$. (d) Similar to (a) but with mismatched fields---case (II).
(e) Detailed behavior of $\rho_{0}$ and $\rho_{1}$ with $\mathcal{M}_{\textrm{II}} =0$. Squares correspond to measured values, whereas  solid lines and bars are fits. Dashed lines show the expected behavior for an ideal single-photon Fock state when [(b) and (c)] $\mathcal{M}=1$ and (e) $\mathcal{M}=0$.}
\label{fig_displaced}
\end{figure}

First,  we carefully adjusted the overlap between signal and reference and recorded the statistics of the displaced single-photon state---case (I). The measured statistics for several values of displacement are shown in Fig.~\ref{fig_displaced}(a). Ideally, the vacuum component $\rho_{0}$ of a weakly displaced single-photon state grows as a function of $|\alpha|$, whereas the one-photon component $\rho_{1}$ rapidly decreases.  
In Figs~\ref{fig_displaced}(b) and \ref{fig_displaced}(c) we show in more detail the evolution of photon-number contributions $\rho_{0}$--$\rho_{4}$ as a function of the displacement.
We model our results with the help of Eq.~(\ref{eq_overlap}) taking into account a small fraction of higher photon-number contributions in the signal state $\vec{\rho} = [0.002(1), 0.942(2), 0.054(2), 0.002(1)]$.  We convolve $\rho_{n}^{D}$ with the Poissonian term of the mismatched signal and consider the overlap factor as a free parameter. By fitting against vacuum and one-photon components we find $\mathcal{M}_{\textrm{I}} = 0.70(2)$, which is used to predict the behavior of higher photon-number components, and which also is in agreement with our previous investigations.
 
In order to validate our findings with   the  optimized overlap, we next misadjusted the temporal delay  resulting in a joint state of signal and reference---case (II). 
The reconstructed statistics present striking differences in comparison to the previous case. The measured vacuum components remain zero regardless  of the displacement due to the vanishing overlap ($\mathcal{M}_{\textrm{II}}  = 0$) between signal and reference [Fig.~\ref{fig_displaced}(d)]. Moreover, the decrement of the one-photon component is less pronounced than before [Fig.~\ref{fig_displaced}(e)]. The small deviation from the expected behavior  occurs due to the higher photon-number contributions of the signal. Overall, the oscillatory behavior of the photon-number distribution is absent.
 
Even though  the fluctuations of the click statistics are negligible,
the uncertainties  in the inverted statistics can become large due to the low detection efficiency and the limited amount of collected statistics. To evaluate the error bars of the reconstructed statistics we employ  a Monte Carlo simulation. Assuming normal distributions, we attribute to  
 each  component of the recorded clicks a standard deviation given by its square root. More than a thousand of such simulations, compatible to our experimental data, were realized and we kept the ones respecting the constraint of non-negative probabilities. The error bars in Fig.~(\ref{fig_displaced}) are given as the deviation from the most probable value. We observe that the errors in the vacuum and one-photon components are negligible, 
 whereas the errors in the higher photon-number components increase faster with respect to $|\alpha|$.

\begin{figure}
\includegraphics[width = 0.48\textwidth]{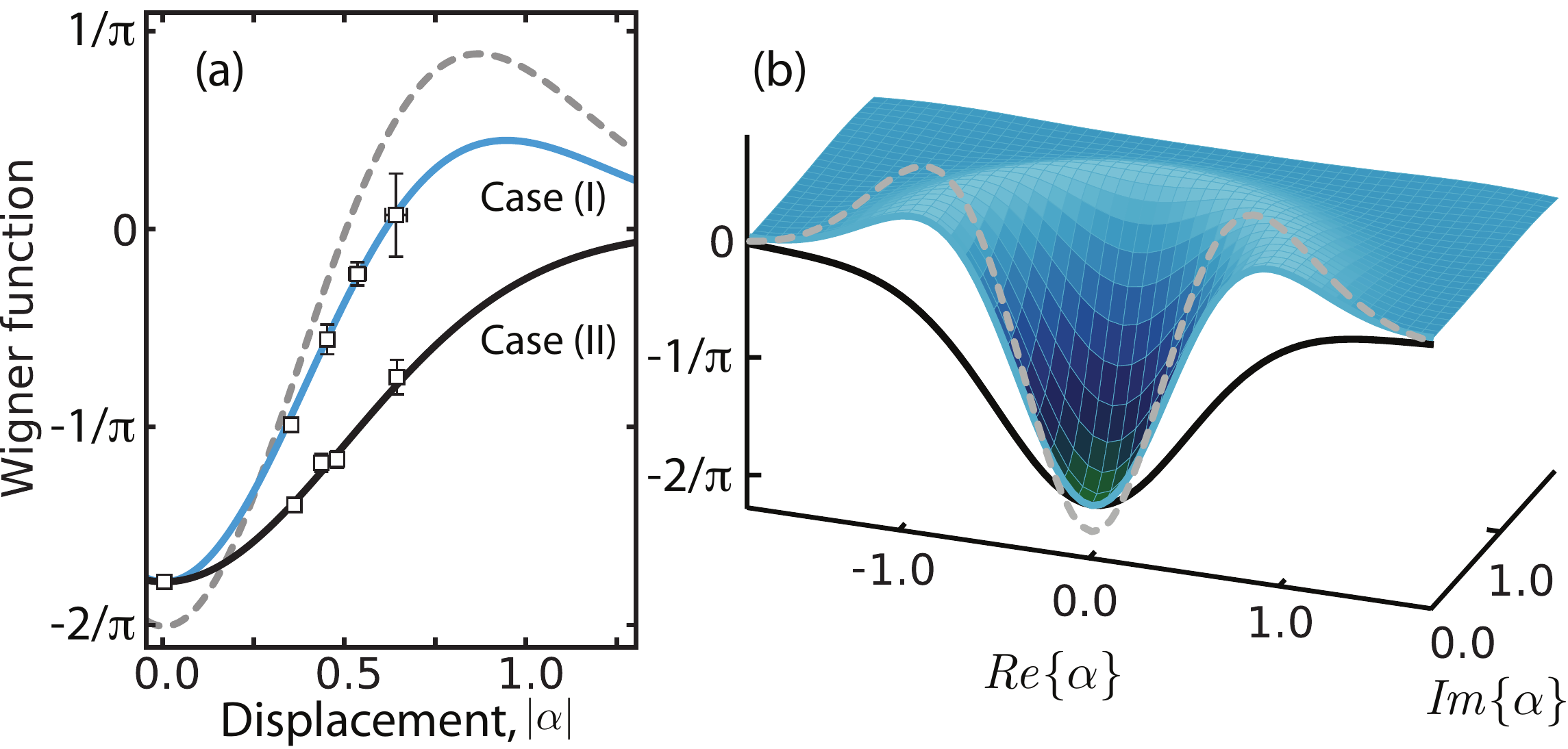}
\caption{(a) Probed values of the Wigner function  as a function of   displacement $|\alpha|$ in cases (I) and (II). (b) Extracted phase-averaged Wigner function in case (I). Squares correspond to probed values, whereas solid lines illustrate the theoretical model. The dashed line shows the Wigner function of an ideal single-photon Fock state.}
\label{fig_Wigner_function}
\end{figure}

By estimating the expectation value of the parity from the statistics in Figs~\ref{fig_displaced}(a) and \ref{fig_displaced}(d) we probe the phase-averaged Wigner function around the origin as shown in Fig.~\ref{fig_Wigner_function}. We obtain a maximal negativity of $-0.565(4)$ [Fig.~\ref{fig_Wigner_function}(a)], which clearly signalizes the nonclassicality of the single-photon state. This value deviates only slightly from the one expected for a single-photon Fock state ($-0.637$), and can be completely explained by the presence of higher photon-number contributions. The extracted values of the Wigner function in case (I) [Fig.~\ref{fig_Wigner_function}(b)] follow in good agreement the expected behavior. The observed broadening at larger values of displacement reflects the slight mode mismatch~\cite{K.Laiho2009}. In case (II) our detection scheme can observe the nonclassicality of the joint state even with increasing Poissonian term. In conclusion, by applying a properly prepared reference field, our detection technique is not only restricted to the photon-number degree of freedom but  can also unravel the single- or multimode characteristics of the state in other degrees of freedom.

 To summarize, we have directly  probed the Wigner function of a nonclassical single-photon wave packet in a loss-tolerant fashion. Our detector can verify the nonclassicality of the state and also highlight the role of mode properties in the detection. Thus, it can become very attractive for addressing the quantum information contained in individual spectral modes. The point-by-point capability---related to the connection of the parity with the $\delta$ function in phase space---allows us  to concentrate on the interesting regions of the Wigner function and can open new routes for the characterization of quantum optical states with time-multiplexed detection \cite{Vogel2005}. 
 
This work was supported by the EC under the grant agreement CORNER (FP7-ICT-213681). KNC~acknowledges  the support from the AvH  Foundation.

 


\begin{thebibliography}{10}

\bibitem{Lvovsky2009}
A.~I.~Lvovsky and M.~G. Raymer,  Rev.~Mod.~Phys.~\textbf{81}, 299 (2009),   and the references therein.


\bibitem{A.I.Lvovsky2001}
A. I. Lvovsky {\it et al.},  Phys. Rev. Lett. \textbf{87}, 050402 (2001);
A. Zavatta, S.~Viciani, and M.~Bellini, Science  \textbf{306}, 660 (2004);
A. Ourjoumtsev {\it et al.},   {\it ibid.} \textbf{312}, 83 (2006);
J. S. Neergaard-Nielsen {\it et al.}, Phys. Rev. Lett. \textbf{97}, 083604 (2006);
 K. Wakui {\it et al.}, Opt. Express \textbf{15}, 3568 (2007);
T. Gerrits {\it et al.},  Phys. Rev. A  \textbf{82}, 031802(R) (2010).

\bibitem{Luetkenhaus95}
N. L\"utkenhaus and S. M. Barnett,
Phys. Rev. A \textbf{51}, 3340 (1995).
  

\bibitem{Englert1993}
K. E. Cahill and R. J. Glauber, Phys. Rev. \textbf{177}, 1882 (1969);
A. Royer, Phys. Rev. A \textbf{15}, 449 (1977); Phys. Rev. Lett. \textbf{55}, 2745 (1985);
B.-G. Englert, N. Sterpi, and H. Walther,  Optics Comm. \textbf{100}, 526 (1993).

\bibitem{Leibfried1996}
 D.~Leibfried {\it et al.}, ~Phys.~Rev.~Lett.~\textbf{77}, 4281 (1996).

\bibitem{K.Banaszek1996}
K.~Banaszek and K.~Wodkiewicz, Phys.~Rev.~Lett.~\textbf{76}, 4344 (1996);
S.~Wallentowitz and W.~Vogel, Phys.~Rev.~A \textbf{53}, 4528 (1996).

\bibitem{P.Bertet2002}
P.~Bertet  {\it et al.}, Phys.~Rev.~Lett.~\textbf{89}, 200402 (2002).

\bibitem{K.Banaszek1999}
K. Banaszek {\it et al.}, Phys. Rev. A \textbf{60}, 674  (1999).

\bibitem{Bondani2009}
M. Bondani, A. Allevi, and A. Andreoni, Opt. Lett. \textbf{34}, 1444 (2009);
A. Allevi  {\it et al.},  Phys. Rev. A \textbf{80}, 022114 (2009).

\bibitem{Banaszek1998}
K. Banaszek and K. Wodkiewicz, Phys. Rev. A  \textbf{58}, 4345 (1998);
A. Kuzmich, I. A. Walmsley, and L. Mandel,  Phys. Rev. Lett. \textbf{85}, 1349 (2000).

\bibitem{J.Eisert2002}
J. Eisert, S. Scheel, and M. B. Plenio, Phys. Rev. Lett. \textbf{89}, 137903 (2002); 
J. Fiurasek, {\it ibid.} \textbf{89}, 137904 (2002);
G. Giedke and J. I. Cirac, Phys. Rev. A \textbf{66} 032316 (2002).

\bibitem{D.Achilles2003}
D. Achilles {\it et al.}, Opt. Lett. \textbf{28}, 2387 (2003);
M. J. Fitch {\it et al.}, Phys. Rev. A \textbf{68}, 043814 (2003).

\bibitem{D.Achilles2005}
D. Achilles, C. Silberhorn, and I. A. Walmsley, Phys. Rev. Lett. \textbf{97}, 043602 (2006);
M. Avenhaus {\it et al.},  {\it ibid.} \textbf{101}, 053601 (2008).

\bibitem{Lundeen2009}
J. S. Lundeen {\it et al.}, Nature Physics \textbf{5}, 27 (2009).

\bibitem{Puentes2009}
G. Puentes  {\it et al.}, Phys. Rev. Lett. \textbf{102}, 080404 (2009).
  
\bibitem{Avenhaus2010}
M.~Avenhaus  {\it et al.},  Phys. Rev. Lett. \textbf{104}, 063602 (2010).


\bibitem{Wittmann2010}
C.~Wittmann {\it et al.},  Phys. Rev. Lett. \textbf{104}, 100505 (2010).


\bibitem{W.Schleich1987}
W.~Schleich and J.~A.~Wheeler, Nature  \textbf{326}, 574 (1987).

\bibitem{Oliveira1990}
F. A. M. de Oliveira {\it et al.},  Phys. Rev. A \textbf{41}, 2645 (1990);
A. I. Lvovsky and S. A. Babichev, {\it ibid.} \textbf{66},  011801(R) (2002). 

\bibitem{Banaszek2002}
K. Banaszek  {\it et al.},  Phys. Rev. A \textbf{66}, 043803 (2002).
 
 \bibitem{K.Laiho2009}
K. Laiho {\it et al.},  New. J. Phys. \textbf{11}, 043012 (2009).


\bibitem{D.Achilles2004}
D. Achilles, {\it et al.},   J. Mod. Opt. \textbf{51}, 1499 (2004).


\bibitem{Laiho2009}
K.~Laiho, K.~N.~Cassemiro, and C.~Silberhorn, Opt.~Express \textbf{17},  22823 (2009);
K.~N.~Cassemiro, K.~Laiho, and C.~Silberhorn,   New. J. Phys. \textbf{12}, 113052 (2010).

\bibitem{Vogel2005}
E. V. Shchukin and W. Vogel, Phys. Rev. A \textbf{72}, 043808 (2005).

\end{thebibliography}
\end{document}